\documentclass[12pt,english]{article}
\usepackage[T1]{fontenc}
\usepackage[latin1]{inputenc}
\usepackage{a4wide}
\usepackage{graphicx}

\makeatletter

\usepackage{babel}
\makeatother
\begin{document}

\title{Entangled spin clusters : some special features}

\author{Amit Tribedi and Indrani Bose }

\maketitle
\begin{center}Department of Physics\end{center}

\begin{center}Bose Institute\end{center}

\begin{center}93/1, Acharya Prafulla Chandra Road\end{center}

\begin{center}Kolkata - 700 009, India\end{center}

\begin{abstract}
In this paper, we study three specific aspects of entanglement in
small spin clusters. We first study the effect of inhomogeneous exchange
coupling strengths on the entanglement properties of the S=$\frac{1}{2}$
antiferromagnetic linear chain tetramer compound $NaCuAsO_{4}$. The
entanglement gap temperature, $T_{E}$, is found to have a non-monotonic
dependence on the value of $\alpha$, the exchange coupling inhomogeneity
parameter. We next determine the variation of $T_{E}$ as a function
of $S$ for a spin dimer, a trimer and a tetrahedron. The temperature
$T_{E}$ is found to increase as a function of $S$ but the scaled
entanglement gap temperature $t_{E}$ goes to zero as $S$ becomes
large. Lastly, we study a spin-1 dimer compound to illustrate the
quantum complementarity relation. We show that in the experimentally
realizable parameter region, magnetization and entanglement plateaus
appear simultaneously at low temperatures as a function of the magnetic
field. Also, the sharp increase in one quantity as a function of the
magnetic field is accompanied by a sharp decrease in the other so
that the quantum complementarity relation is not violated.
\end{abstract}

\section*{I. INTRODUCTION}

Entanglement is a key feature of quantum mechanical systems and gives
rise to non-local correlations over and above those expected from
classical considerations \cite{key-1}. It can be of different types
: bipartite, multipartite, zero-temperature, finite-temperature etc.
for which suitable measures are available in certain cases \cite{key-2,key-3,key-4,key-5,key-6,key-7,key-8}.
In the past few years, quantum spin systems have been extensively
studied to gain knowledge on the different aspects of entanglement.
The spins in such systems interact via the exchange interaction and
also with an external field, if any. Several studies show that the
amount of entanglement can be changed by varying the temperature T
and/or the magnitude of the external field \cite{key-3,key-4,key-9}.
In the case of entangled thermal states, one can define a critical
temperature below which entanglement is present in the system and
above which entanglement vanishes, i.e., the system becomes separable.
Detection of entanglement can be made with the help of an entanglement
witness (EW) which is an observable the expectation value of which
is positive in separable and negative in entangled states \cite{key-10,key-11,key-12}.
Thermodynamic observables like internal energy, magnetization and
susceptibility have been proposed as EWs \cite{key-12,key-13,key-14}.
There is now experimental evidence that entanglement can affect the
macroscopic properties of solids like specific heat and magnetic susceptibility
\cite{key-15}. Recently, it has been shown that for separable states,
the sum of magnetic susceptibilities in the three orthogonal directions
x, y, and z obeys the inequality

\begin{equation}
\bar{\chi}\equiv\chi_{x}+\chi_{y}+\chi_{z}\geq\frac{N\, S}{k_{B}T}\label{1}\end{equation}

\noindent where $N$ is the total number of spins in the system, $S$
the magnitude of the spin, $k_{B}$ the Boltzmann constant and $T$
the temperature \cite{key-14}. If the magnetization operator $M_{\alpha}=\sum_{j}S_{j}^{\alpha}$
commutes with the Hamiltonian H of the system, i.e., $[H,M_{\alpha}]=0$,
the magnetic susceptibility $\chi_{\alpha}$ ($\alpha=$x, y, z) can
be written as 

\begin{equation}
\chi_{\alpha}=\frac{1}{k_{B}T}\left[\left\langle \left(M_{\alpha}\right)^{2}\right\rangle -\left\langle M_{\alpha}\right\rangle ^{2}\right]\label{2}\end{equation}

\begin{equation}
=\frac{1}{k_{B}T}\left[\sum_{i,j=1}^{N}\left\langle S_{i}^{\alpha}S_{j}^{\alpha}\right\rangle -\left\langle \sum_{i=1}^{N}S_{i}^{\alpha}\right\rangle ^{2}\right]\label{3}\end{equation}

\noindent Thermodynamic properties in general relate to macroscopic
systems and the thermal state of such a system is entangled if $\bar{\chi}$
is $<$ $\frac{N\, S}{k_{B}T}$ (Eq. (1)). Using the susceptibility
inequality as an EW, one can detect entanglement from the experimental
data without requiring a knowledge of the Hamiltonian of the system.

For a multipartite Hamiltonian $H$, one can define the entanglement
gap as 

\begin{equation}
G_{E}=E_{sep}-E_{g}\label{4}\end{equation}

\noindent where $E_{sep}$ is the minimum separable energy and $E_{g}$
the ground state energy of the Hamiltonian \cite{key-12}. For a spin
Hamiltonian, $E_{sep}$ is the ground state energy of the equivalent
classical Hamiltonian \cite{key-11}. If a system has entanglement
gap $G_{E}>0$, then one can define the entanglement gap temperature,
$T_{E}$, as the temperature at which the thermal (internal) energy
$U(T_{E})=E_{sep}$. For temperature $T<T_{E}$, the thermal state
of the system is bound to be entangled. Recently, a quantum complementarity
relation has been proposed \cite{key-14} between the thermodynamic
observables, magnetization and magnetic susceptibility. This is given
by

\begin{equation}
1-\frac{k_{B}T\bar{\chi}}{N\, S}+\frac{\left\langle \overrightarrow{M}\right\rangle ^{2}}{N^{2}S^{2}}\leq1\label{5}\end{equation}

\noindent where $\left\langle \overrightarrow{M}\right\rangle ^{2}\equiv\left\langle M_{x}\right\rangle ^{2}+\left\langle M_{y}\right\rangle ^{2}+\left\langle M_{z}\right\rangle ^{2}$.
Define the quantities\begin{equation}
P=\frac{\left\langle \overrightarrow{M}\right\rangle ^{2}}{N^{2}S^{2}}\;,\; Q=1-\frac{k_{B}T\bar{\chi}}{N\, S}\label{6}\end{equation}

\noindent The quantity $P$, which depends upon the magnetization,
describes the local properties of individual spins whereas $Q$, which
involves the susceptibility, is representative of quantum spin-spin
correlations. From Eq. (1), a nonzero positive value of $Q$ implies
the presence of entanglement in the system, i.e., non-local correlations.
The complementarity relation shows that the non-local properties are
enhanced at the expense of the local properties in order that $P+Q$
is $\leq1$.

The EWs based on the internal energy and susceptibility have been
used to study the entanglement properties of the spin-$\frac{1}{2}$
antiferromagnetic (AFM) compounds $Cu(NO_{3})_{2},2.5D_{2}$

\noindent $O(CN)$(system of weakly coupled spin dimers) \cite{key-13},
$\left(NHEt\right)_{3}\left[V_{8}^{IV}V_{4}^{V}As_{8}O_{40}\left(H_{2}O\right)\right].H_{2}O$
(system of weakly coupled spin tetramers) \cite{key-16} and the nanotubular
system $Na_{2}V_{3}O_{7}$ (consists of weakly-coupled nine-spin rings)
\cite{key-17}. The weak coupling between the spin clusters allows
each system to be treated as consisting of effectively independent
clusters. Since the clusters contain a few spins, the theoretical
calculation of entanglement-related quantities becomes possible. A
number of molecular magnets are known which are well-described in
terms of small spin clusters such as dimers, trimers, tetramers, tetrahedra
etc \cite{key-18}. For non-bipartite clusters with {}``all-to-all''
spin couplings (trimers, tetrahedra), the EWs based on the internal
energy and susceptibility give the same estimate of the temperature
above which entanglement vanishes \cite{key-16}. For bipartite clusters
(a tetramer describing a square plaquette of spins with only nearest-neighbour(NN)
exchange couplings provides an example), the EW based on the internal
energy can detect only the bipartite entanglement between two qubits
\cite{key-12}. The spin clusters considered so far are described
by Hamiltonians with homogeneous exchange interaction strengths. In
this paper, we consider the $S=\frac{1}{2}$ AFM linear tetramer compound
$NaCuAsO_{4}$ \cite{key-19} in which the linear tetramer consisting
of four spins is described by the Heisenberg Hamiltonian

\begin{equation}
H_{LT}=J\,\overrightarrow{S_{1}}.\overrightarrow{S_{2}}+\alpha J\,\overrightarrow{S_{2}}.\overrightarrow{S_{3}}+J\,\overrightarrow{S_{3}}.\overrightarrow{S_{4}}\label{7}\end{equation}

\noindent We study the entanglement properties of this compound using
both the internal energy and the susceptibility as EWs. We next determine
the entanglement gap temperature $T_{E}$ of small spin clusters as
a function of the magnitude $S$ of spins. Lastly, we determine the
quantities $P$ and $Q$ (Eq. (6)) appearing in the complementarity
relation (Eq. (5)) for a spin-1 dimer compound $[Ni_{2}\,(Medpt)_{2}(\mu-ox)(H_{2}O)_{2}](ClO_{4})_{2}.2H_{2}O$
\cite{key-20} and show that the sharp changes in the magnetization
and the formation of plateaus at low temperatures are accompanied
by sharp changes and plateaus in the amount of entanglement. Magnetization
plateaus have been observed experimentally in the spin-1 dimer compound.
This compound thus provides a concrete example of a system in which
the amount of entanglement can change steeply as a function of the
magnetic field or does not change over a range of field values.

\begin{figure}
\begin{center}\includegraphics{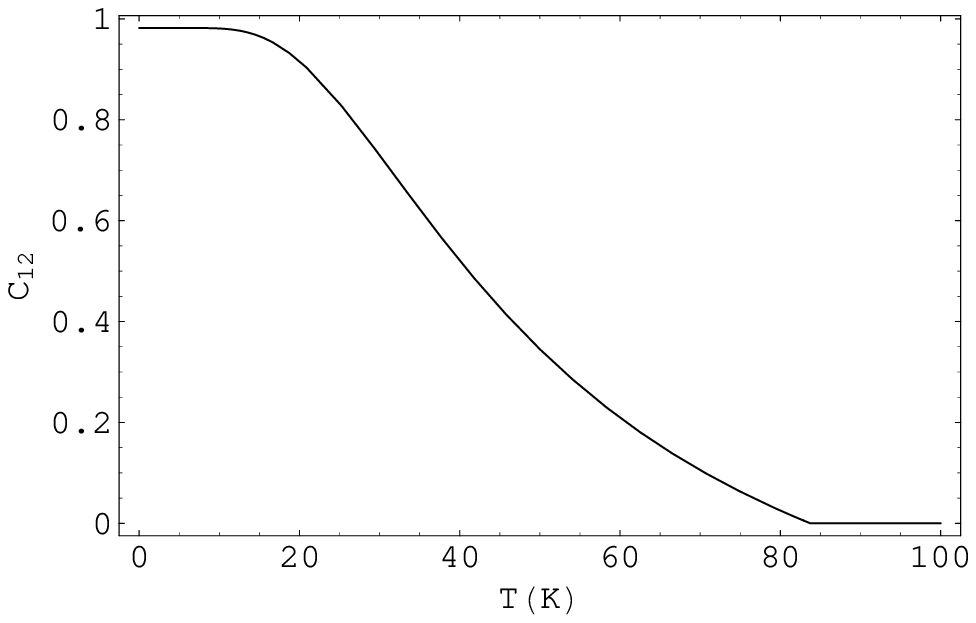}\end{center}

\textbf{FIG. 1.} Concurrence $C_{12}$ as a function of temperature
for $\alpha=0.4$ and $\frac{J}{k_{b}}=92.7K$.
\end{figure}

\begin{figure}
\begin{center}\includegraphics{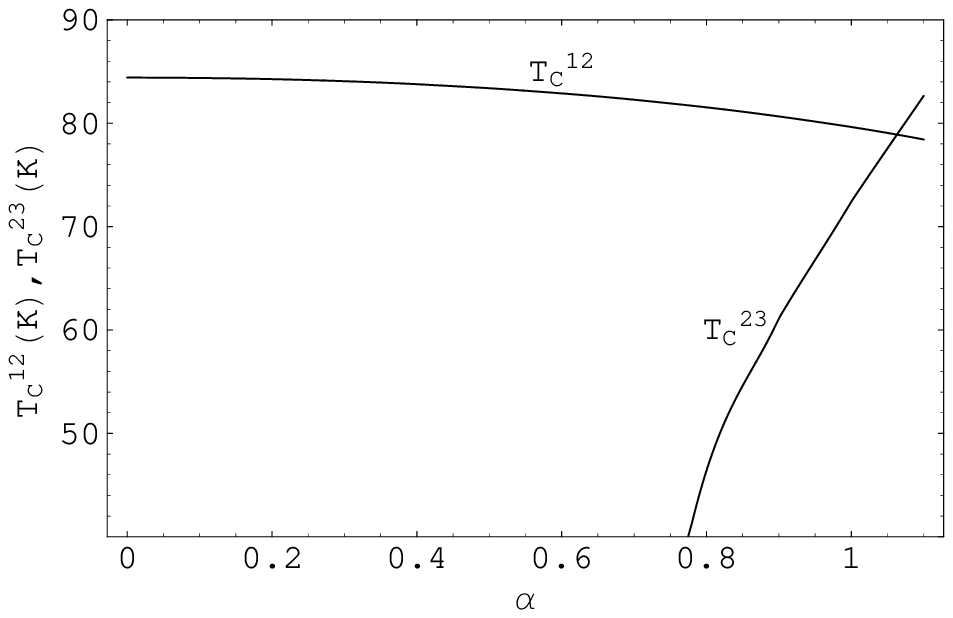}\end{center}

\textbf{FIG. 2.} Plots of $T_{C}^{12}$ and $T_{C}^{23}$ versus $\alpha$.
$T_{C}^{kl}$ is the critical entanglement temperature for the pair
of spins at sites $k$ and $l$.
\end{figure}

\section*{II. LINEAR CHAIN TETRAMER}

The $S=\frac{1}{2}$ AFM compound $NaCuAsO_{4}$ has a linear chain
tetrameric structure described by the Hamiltonian, $H_{LT}$, in Eq.
(7) with $\alpha\approx0.4$. The term {}``linear'' refers to the
pattern of exchange couplings and not to the spatial structure of
the tetramer \cite{key-19}. The total spin $S^{tot}$ of the tetramer
has the values 2, 1 and 0. There are five $S^{tot}=2$ states, nine
$S^{tot}=1$ states and two $S^{tot}=0$ states. The different eigenvalues
and eigenvectors are displayed in Appendix A. We now discuss the finite-temperature
entanglement properties of the linear chain tetramer. The thermal
density matrix $\rho\left(T\right)$ is given by

\begin{equation}
\rho\left(T\right)=\frac{1}{Z}\:\sum_{E_{i}}\sum_{m}e^{-\beta E_{i}}\left|\psi_{i,m}\right\rangle \left\langle \psi_{i,m}\right|\label{8}\end{equation}

\noindent The first summation is over all the independent energy eigenstates
and the second summation includes terms corresponding to the $(2S^{tot}+1)$
degenerate eigenstates with the eigenvalue $E_{i}$. $Z$ denotes
the partition function

\begin{equation}
Z=\sum_{E_{i}}\left(2S^{tot}+1\right)e^{-\beta E_{i}}\label{9}\end{equation}

\noindent A measure of entanglement between the spins at sites $k$
and $l$ is given by the concurrence $C_{kl}$. This is calculated
from the reduced thermal density matrix $\rho_{kl}\left(T\right)$
using standard procedure \cite{key-2,key-9}. Since the eigenvalues
and eigenvectors of the linear chain tetramer are known, the calculation
of $C_{kl}$ is straightforward. Figure 1 shows the variation of $C_{12}$
as a function of temperature for $\alpha=0.4$ and $\frac{J}{k_{B}}=92.7\, K$,
the parameter values relevant for $NaCuAsO_{4}$ \cite{key-18,key-19}.
The concurrence $C_{23}$ is zero for these parameter values. One
can further define a critical temperature $T_{C}^{kl}$ above which
the entanglement between the spins at the sites $k$ and $l$ disappears.
Figure 2 shows a plot of $T_{C}^{12}$ and $T_{C}^{23}$ versus $\alpha$
for $\frac{J}{k_{B}}=92.7K$ . We next calculate the entanglement
gap temperature $T_{E}$ at which the internal energy 

\begin{equation}
U\left(T_{E}\right)=-\frac{1}{Z}\left(\frac{\partial Z}{\partial\beta}\right)=E_{sep}\label{10}\end{equation}

\noindent where $E_{sep}$ is the minimum separable energy. Figure
3 shows a plot of $T_{E}$ versus $\alpha$ for $\frac{J}{k_{B}}=92.7K$
(Curve a). The operator $H-E_{sep}$ is an EW since $Tr\left[\rho\left(H-E_{sep}\right)\right]=U\left(T\right)-E_{sep}$
is $<0$ ($\geq0$) when the thermal state is entangled (separable).
If the Hamiltonian $H$ of the system contains only local interactions
such that $H=\sum_{<ij>}H_{ij}$ and the underlying lattice is bipartite,
then

\begin{equation}
H-E_{sep}=\sum_{<ij>}\left(H_{ij}-e_{sep,ij}\right)\label{11}\end{equation}

\noindent where $e_{sep,ij}$ is the minimum separable energy associated
with the interaction between the spins located at the sites $i$ and
$j$. In the case of a translationally invariant Hamiltonian, $H_{ij}$
and $e_{sep,ij}$ are the same for each interacting spin pair. Each
term in the sum on the RHS of Eq. (11) can be considered as a bipartite
EW. Thus the expectation value of $H-E_{sep}$ is negative only if
the two spins in the interacting spin pairs are entangled. In the
case of the linear chain tetramer, the Hamiltonian $H_{LT}$ is not
translationally invariant. This is true even in the limit $\alpha=1$.
Since, the $H_{ij}$'s and $e_{sep,ij}$'s are no longer the same
for each interaction bond, $T_{C}^{12}=T_{C}^{34}\neq T_{C}^{23}$.
The expectation value of $H-E_{sep}$ now depends on the relative
magnitudes and signs of the two types of terms on the RHS of Eq. (11).
In contrast, consider a closed chain of four spins in which the NN
spins interact with the same exchange interaction strength. In this
case, because of translational invariance, $T_{C}^{12}=T_{C}^{23}=T_{C}^{34}=T_{C}^{41}=T_{C}$
and the entanglement gap temperature $T_{E}$ is equal to $T_{C}$,
the critical temperature beyond which the entanglement between two
NN spins vanishes (concurrence is zero). In the case of the linear
chain tetramer, a similar interpretation cannot be given.

We now use the magnetic susceptibility as an EW to determine the critical
temperature $T_{C}^{\chi}$ beyond which the thermal state of the
linear chain tetramer is separable. We consider the case of zero-field
susceptibility. In the absence of a magnetic field, $\left\langle M_{\alpha}\right\rangle =0\,(\alpha=x,y,z)$.
Also, due to the spin isotropy of the Hamiltonian, $H_{LT}$, ($S^{tot}$
is a good quantum number), $\chi_{x}=\chi_{y}=\chi_{z}=\chi$. The
susceptibility $\chi$ can be written as

\begin{equation}
\chi=\frac{\beta}{3Z}\:\sum_{E_{i}}\left(2S^{tot}+1\right)\left(S^{tot}+1\right)S^{tot}e^{-\beta E_{i}}\label{12}\end{equation}

\noindent The susceptibility inequality for separable states (Eq.
(1)) becomes

\begin{equation}
\chi\geq\frac{NS}{3k_{B}T}\label{13}\end{equation}

\noindent The critical temperature $T_{C}^{\chi}$ is given by the
intersection point of the two curves : $\chi$ versus $T$ plot from
Eq. (12) and $\chi$ versus $T$ plot from the equality in Eq. (13)
\cite{key-14,key-16}. For $\alpha=0.4$ and $\frac{J}{k_{B}}=92.7\, K$,
one obtains the estimate $T_{C}^{\chi}=90.88\, K$, which is really
the lower bound of the critical temperature above which entanglement
vanishes. Figure 3 also shows the variation of $T_{C}^{\chi}$ as
a function of $\alpha$ (Curve b). Since the linear chain tetramer
is associated with a bipartite graph, $T_{C}^{\chi}$ is $>T_{E}$,
the entanglement gap temperature.

\begin{figure}
\begin{center}\includegraphics{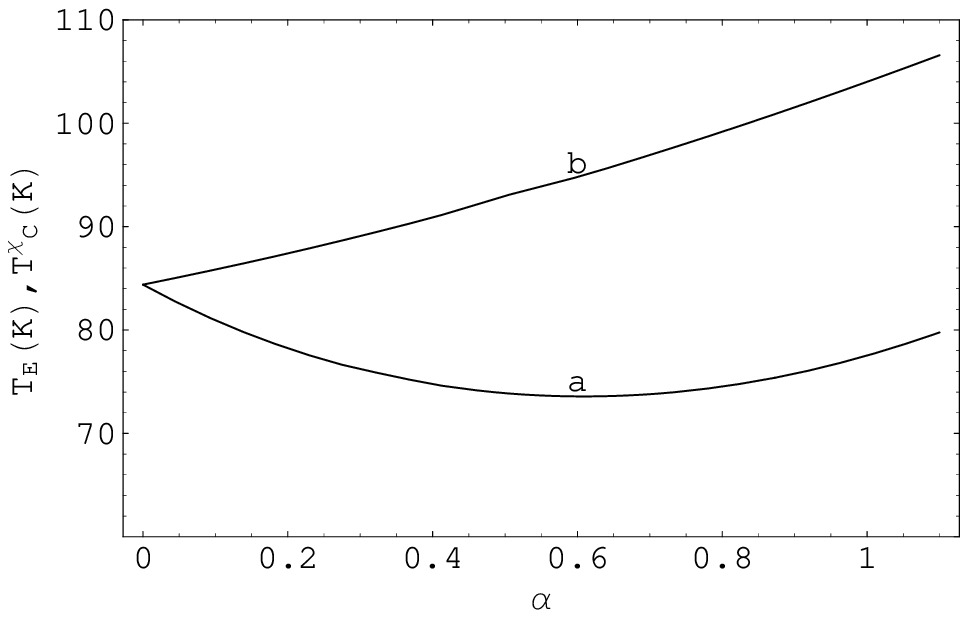}\end{center}

\textbf{FIG. 3.} Plots of $T_{E}$ (Curve a) and $T_{C}^{\chi}$ (Curve
b) for the linear chain tetramer as a function of $\alpha$ ($\frac{J}{k_{B}}=92.7\, K$). 
\end{figure}

\section*{III. GENERAL SPIN $S$ AND $T_{E}$}

We have so far considered the case $S=\frac{1}{2}$. We now consider
dimers, trimers and tetrahedra of spins of magnitude $S$. The Hamiltonians
describing the small clusters are 

\begin{equation}
H_{dimer}=J\,\overrightarrow{S_{1}}.\overrightarrow{S_{2}}\label{14}\end{equation}

\begin{equation}
H_{trimer}=J\,\left(\overrightarrow{S_{1}}.\overrightarrow{S_{2}}+\overrightarrow{S_{2}}.\overrightarrow{S_{3}}+\overrightarrow{S_{3}}.\overrightarrow{S_{1}}\right)\label{15}\end{equation}

\begin{equation}
H_{tetrahedron}=J\,\left(\overrightarrow{S_{1}}.\overrightarrow{S_{2}}+\overrightarrow{S_{2}}.\overrightarrow{S_{3}}+\overrightarrow{S_{3}}.\overrightarrow{S_{4}}+\overrightarrow{S_{4}}.\overrightarrow{S_{1}}+\overrightarrow{S_{1}}.\overrightarrow{S_{3}}+\overrightarrow{S_{2}}.\overrightarrow{S_{4}}\right)\label{16}\end{equation}

\noindent A trimer and a tetrahedron are defined on a non-bipartite
graph. The AFM cluster Hamiltonian in each case contains {}``all-to-all''
spin couplings and is frustrated as there is no separable state that
simultaneously minimizes the energy of each interacting spin pair.
Thus $e_{sep,ij}$ in Eq. (11) is no longer the minimum separable
energy of an interacting spin pair, it has a magnitude greater than
that of the latter quantity. The minimum separable energy for the
whole Hamiltonian is $E_{sep}=\sum_{<ij>}e_{sep,ij}=N_{tot}\, e_{sep}$,
where $N_{tot}$ is the total number of interacting spin pairs. Since
$e_{sep}$ is greater than the minimum separable energy for an interacting
spin pair, the EW, $H-E_{sep}$, can detect entanglement even if the
entanglement between the spins in the interacting spin pair vanishes,
i.e., the corresponding reduced density matrix becomes separable \cite{key-12}.
In this case, the entanglement gap temperature $T_{E}$ is $T_{C}$,
the critical temperature beyond which the NN concurrence is zero.
As shown in \cite{key-16}, $T_{E}=T_{C}^{\chi}$ for $S=\frac{1}{2}$
non-bipartite clusters like the trimer and the tetrahedron described
by the Heisenberg Hamiltonian with homogenous exchange couplings.
This result holds true for general $S$ in the case of spin clusters
with {}``all-to-all'' homogeneous Heisenberg spin couplings. We
thus use only the internal energy-based EW to determine how the critical
entanglement temperature varies as a function of $S$ in the cases
of the spin dimer, trimer and the tetrahedron.

For Hamiltonians with {}``all-to-all'' spin couplings, the energy
eigenvalues of all the eigenstates can be determined quite easily
from a simple formula. The Hamiltonian can be written as

\begin{equation}
H=\frac{1}{2}\left[\left(\overrightarrow{S}^{tot}\right)^{2}-\sum_{i=1}^{N}S_{i}^{2}\right]\label{17}\end{equation}

\noindent where $\overrightarrow{S}^{tot}=\sum_{i=1}^{N}\overrightarrow{S}_{i}$.
The eigenvalue $E_{S^{tot}}$ for a state with total spin $S^{tot}$
is

\begin{equation}
E_{S^{tot}}=\frac{1}{2}\left[S^{tot}\left(S^{tot}+1\right)-NS\left(S+1\right)\right]\label{18}\end{equation}

\noindent where $S$ is the magnitude of a spin. The possible values
of $S^{tot}$ are $NS$, $NS-1$, .......etc. The lowest value is
zero for $N$ even and $\frac{1}{2}$ for $N$ odd. Under the vector
addition of angular momenta, a particular $S^{tot}$ value can be
achieved in more than one way, i.e., has some multiplicity. Let $P_{S^{tot}N}^{S}$
be the multiplicity, i.e., the number of possible states with total
spin angular momentum $S^{tot}$ when $N$ spins, each of magnitude
$S$, are combined. As shown by Mikhailov \cite{key-21}, $P_{S^{tot}N}^{S}$
is given by

\begin{equation}
P_{S^{tot}N}^{S}=\sum_{k}\left(-1\right)^{k}\left(\begin{array}{c}
N\\
k\end{array}\right)\left(\begin{array}{c}
N(S+1)-S^{tot}-(2S+1)k-2\\
N-2\end{array}\right)\label{19}\end{equation}

\noindent Here $\left(\begin{array}{c}
m\\
n\end{array}\right)$ are the binomial coefficients. The summation index $k$ satisfies
two conditions : $(i)\, k\geq0$ and $(ii)\,$the upper numbers in
the binomial coefficients cannot be less than the lower numbers. Thus,
$0\leq k\leq[\frac{(SN-S^{tot})}{2S+1}]$ where $[b]$ denotes the
integer part of $b$. The minimum separable energy of a spin cluster
is equal to the ground state energy of the equivalent classical Hamiltonian.
In the classical ground state, $S^{tot}=0$ and each $<S_{i}^{2}>=S^{2}$.
Thus, the minimum separable energy, $E_{sep}$, for the dimer, trimer
and tetrahedron is given by $E_{sep}=-S^{2}$(dimer), $-\left(\frac{3}{2}S^{2}\right)$
(trimer) and $-2S^{2}$(tetrahedron). The entanglement gap temperature
$T_{E}$ can be calculated by using the relation in Eq. (10). Figure
4 shows the variation of $T_{E}$ with $S$ for dimers (star), trimers
(solid square) and tetrahedra (solid diamond). The entanglement gap
temperature, $T_{E}$, is found to increase with $S$ in each case.
According to conventional notion, spins behave as classical objects
in the limit of large $S$. The commutation bracket of spin operators,
with each operator scaled by the total spin $S$, tends to zero as
$S\rightarrow\infty$. One would thus expect the entanglement gap
temperature $T_{E}$ to decrease rather than increase as the magnitude
of $S$ is raised. Some earlier studies have reported findings similar
to ours. Hao and Zhu \cite{key-22} have studied the AFM Heisenberg
chain with spins of magnitude $S$. For a two-sited chain, i.e., a
dimer, they find that the entanglement gap temperature $T_{E}$ increases
almost linearly with $S$. For $S=1$, they have shown that $T_{E}$
decreases as the length of the chain is increased. Wie\'{s}niak et
al. \cite{key-14} have determined the critical entanglement temperature,
$T_{C}^{\chi}$, based on the susceptibility as an EW, and find the
result that $T_{C}^{\chi}=1.6\, J$ for the $S=\frac{1}{2}$ Heisenberg
chain and $T_{C}^{\chi}=2\, J$ for a chain of spins 1. As pointed
out by Dowling et al. \cite{key-12}, it is sensible to define a scaled
temperature

\begin{equation}
t=\frac{k_{B}T}{E_{tot}}\label{20}\end{equation}

\noindent for a meaningful comparison of Hamiltonians with different
total energy ranges, $E_{tot}$ ($E_{tot}$ is the difference between
the highest and the lowest energy eigenvalues). The scaled entanglement
gap temperature can be defined as $t_{E}=\frac{k_{B}T_{E}}{E_{tot}}$.
The inset of Fig. 4 shows the variation of $t_{E}$ with $S$ for
dimers (star), trimers (solid triangle) and tetrahedra (solid square).
One finds that the scaled entanglement gap temperature decreases as
$S$ increases. The result can be interpreted in the following way.
As $S$ increases, the fraction of the total energy range of the spin
system which corresponds to entangled states decreases and tends to
a limiting value as $S\rightarrow\infty$. Classical behaviour presumably
emerges when the entangled states have a negligible contribution to
the total energy range.

\begin{figure}
\begin{center}\includegraphics{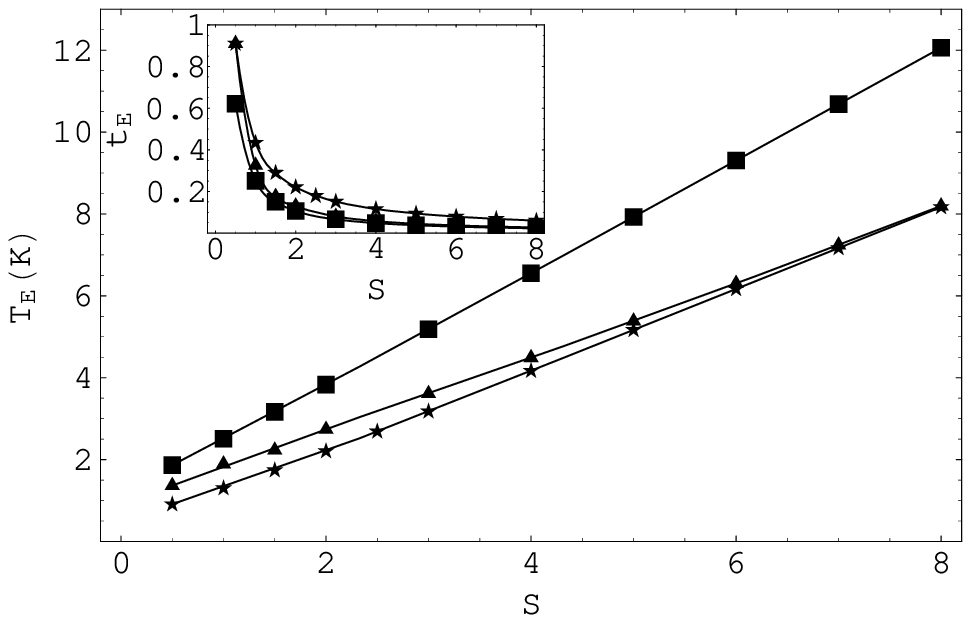}\end{center}

\textbf{FIG. 4.} Variation of the entanglement gap temperature, $T_{E}$,
with $S$ for dimers (star), trimers (solid triangle) and tetrahedra
(solid square). The inset shows the variation of the scaled entanglement
gap temperature $t_{E}$ with $S$.
\end{figure}

\section*{IV. SPIN-1 DIMER : QUANTUM COMPLEMENTARITY}

Bose and Chattopadhyay \cite{key-23} have considered some toy spin
models and shown that first order quantum phase transitions, occurring
at special values of the external magnetic field, are accompanied
by magnetization and entanglement jumps. Upward jumps in the magnetization
give rise to downward jumps in the amount of entanglement. Also, magnetization
and entanglement plateaus coexist in the same range of magnetic fields.
Later studies established the general validity of these results \cite{key-24,key-25,key-26}.
In this section, we show that the quantum complementarity relation
(Eq. (5)) provides a natural explanation for the correlated changes
in the amounts of magnetization and entanglement as a function of
the magnetic field. We illustrate this in the case of a spin-1 dimer
compound $[Ni_{2}\,(Medpt)_{2}(\mu-ox)(H_{2}O)_{2}](ClO_{4})_{2}.2H_{2}O$
($Medpt=methyl-bis(3-aminopropyl)amine$) which exhibits magnetization
plateaus at sufficiently low temperatures \cite{key-20}. The Hamiltonian
describing the spin-1 dimer is

\begin{equation}
H_{d}=J\left(S_{1}^{x}S_{2}^{x}+S_{1}^{y}S_{2}^{y}\right)+\delta J\left(S_{1}^{z}S_{2}^{z}\right)+d\left[\left(S_{1}^{z}\right)^{2}+\left(S_{2}^{z}\right)^{2}\right]+B\left(S_{1}^{z}+S_{2}^{z}\right)\label{21}\end{equation}

\noindent where $\delta$ is the exchange anisotropy parameter, $d$
labels the axial zero-field splitting parameter and $B$ the strength
of the external magnetic field. The negative (positive) sign of the
parameter $d$ corresponds to an easy-axis (easy-plane) single ion
anisotropy. If the spin system is entangled, a sharp increase in the
magnetization (obtained at low temperatures) is accompanied by a sharp
decrease in the amount of entanglement so that the complementarity
relation is not violated. At $T=0$, the sharp changes become the
`jumps' associated with first order quantum phase transitions. As
$T$ increases, the changes occur more gradually as a function of
the magnetic field. In an entangled system, the magnetization plateaus
are accompanied by entanglement plateaus and the complementarity relation
continues to be valid. To illustrate this, we first calculate the
eigenvalues and the eigenvectors of the dimer Hamiltonian $H_{d}$
(Eq. (37)). These are displayed in Appendix B. The magnetization $M$
(only the z-component is non-zero) and $\chi_{z}$, the z-component
of the susceptibility are derived from 

\begin{equation}
M=\frac{1}{\beta Z}\frac{\partial Z}{\partial\beta},\;\chi_{z}=\frac{\partial M}{\partial B}\label{22}\end{equation}

\noindent The susceptibility components $\chi_{x}$ and $\chi_{y}$
are determined from Eq. (3) with $\left\langle S_{1}^{x}\right\rangle ,\,\left\langle S_{1}^{y}\right\rangle ,\,\left\langle S_{2}^{x}\right\rangle $
and $\left\langle S_{2}^{y}\right\rangle =0$ since the magnetic field
is in the z-direction. One can now calculate the terms $P$ and $Q$
(Eq. (6)) appearing in the quantum complementarity relation given
by Eq. (5).

\begin{figure}
\begin{center}\includegraphics{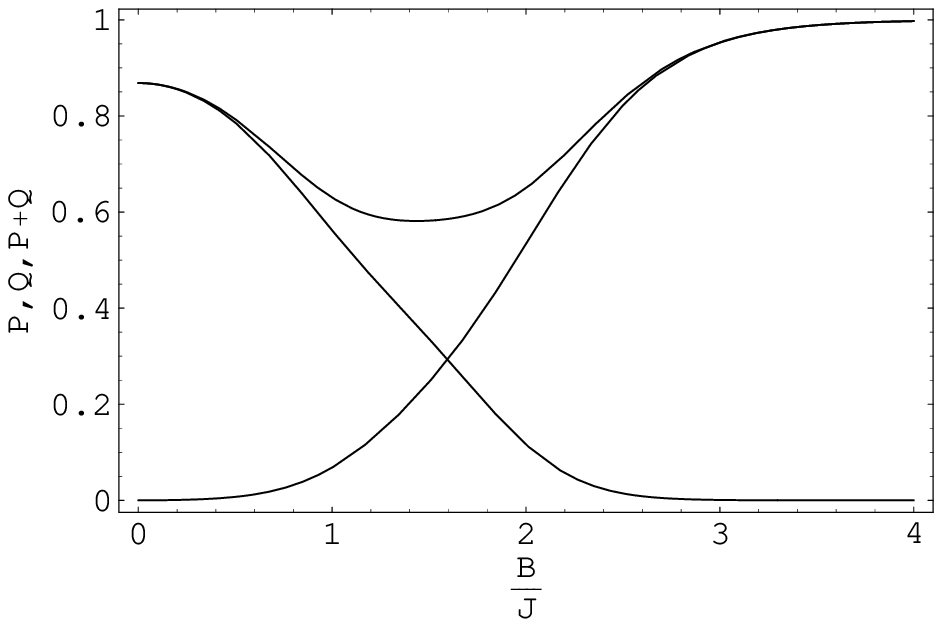}\end{center}

\textbf{FIG. 5.} Plots of $P$, $Q$ and $P+Q$ as a function of $\frac{B}{J}$
in the case of a spin-1 dimer ($\delta=1,\, d=0$ and $\beta J=3$.)
\end{figure}

\begin{figure}
\begin{center}\includegraphics{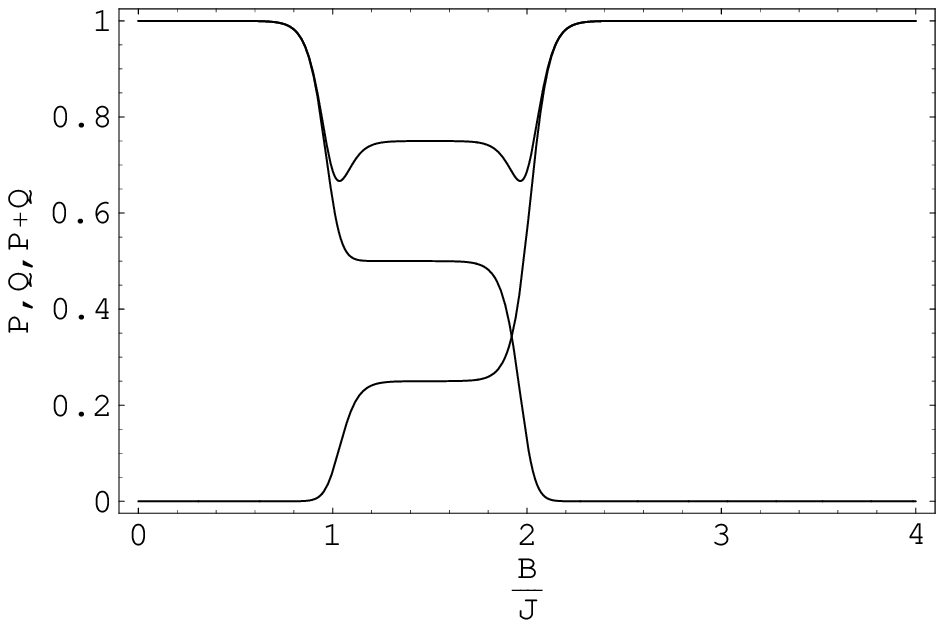}\end{center}

\textbf{FIG. 6.} Plots of $P$, $Q$ and $P+Q$ as a function of $\frac{B}{J}$
in the case of a spin-1 dimer ($\delta=1,\, d=0$ and $\beta J=20$.)
\end{figure}

Figure 5 shows the plots of $P$, $Q$ and $P+Q$ as a function of
$\frac{B}{J}$ for $\delta=1,\, d=0$ and $\beta J=3$. Figure 6 shows
the appearance of plateaus as the temperature is lowered ($\beta J=20$).
Note that a sharp increase in $P$ is accompanied by a sharp decrease
in $Q$. Plateaus in $P$ and $Q$ occur in the same range of magnetic
fields. At sufficiently low temperatures, the two-step plateau structure
is still obtained for non-zero values of $d$. The intermediate plateau
has a lesser width for negative values of $d$ and disappears at $d=-1$.
At this point, $Q$ is $\leq0$ throughout the range of $\frac{B}{J}$
values so that the spin system is not entangled. If $d$ is changed
from $d=-1$ to $d=+1$, the two-step structure in both $P$ and $Q$
is recovered and the amount of entanglement is no longer zero. The
easy-plane single ion anisotropy ($d>0$) is found to be favourable
towards plateau formation in both $P$ and $Q$. The changes in $P$
and $Q$ are correlated so that the complementarity relation $P+Q\leq1$
is always valid. A two-step magnetization curve has been experimentally
observed in the spin-1 nickel compound mentioned earlier \cite{key-20}.
Theoretical calculations, based on a description of the compound as
a collection of independent spin-1 dimers, give a good fit to the
experimental data on the magnetization and susceptibility. The exchange
anisotropy parameter $\delta$ has been taken as 1 and the single
ion anisotropy is of the easy-axis type ($d<0$). Magnetization experiments
have been carried out for the external magnetic field parallel to
the z and x directions. In both the cases, a two-plateau structure
has been seen in the magnetization versus field curves. In the latter
case, the plateau structure is found to be more prominent. Our theoretical
calculations suggest that the magnetization plateaus exhibited by
the spin-1 nickel compound are accompanied by entanglement plateaus.

\section*{V. SUMMARY AND DISCUSSION }

In this paper, we study some special features of entangled small spin
clusters. We first consider the $S=\frac{1}{2}$ AFM linear chain
tetramer compound, $NaCuAsO_{4}$, described by the Heisenberg exchange
interaction Hamiltonian with inhomogeneous exchange coupling strengths
(Eq. (7)) and show that the entanglement gap temperature, $T_{E}$,
has a non-monotonic dependence on the exchange coupling inhomogeneity
parameter $\alpha$. The critical entanglement temperature, $T_{C}^{\chi}$,
obtained by using the susceptibility as an EW, has a monotonic dependence
on $\alpha$. We next determine how the entanglement gap temperature,
$T_{E}$, varies as a function of $S$ in the cases of small spin
clusters like a dimer, a trimer and a tetrahedron. While $T_{E}$
increases with $S$ in each case, the scaled entanglement gap temperature
$t_{E}$ decreases as $S$ increases and goes to zero $S\rightarrow\infty$.
The physical interpretation is that the entangled states have a small
contribution to the total energy range in the limit of large $S$.
The general applicability of this result for spin clusters without
{}``all-to-all'' spin couplings should be investigated. T\'{o}th
\cite{key-11} has considered a Hamiltonian with {}``all-to-all''
couplings between $N$ spin-$\frac{1}{2}$ particles. The entanglement
gap temperature $T_{E}$ is found to increase as $N$ increases but
$t_{E}$ tends to a constant value as $N$ becomes large. Dowling
et al. \cite{key-12} have given examples of Hamiltonians describing
bipartite systems for which $t_{E}$ increases without bound as the
dimension of the Hilbert space associated with the subsystems increases.
In our case, with increasing $S$, the Hilbert space of the system
is enlarged but $t_{E}$ decreases as a function of $S$ and goes
to zero in the limit $S\rightarrow\infty$. This is so since $T_{E}$
has a linear variation with $S$ (Fig. 6) and $E_{tot}$ varies as
$S^{2}$ in the large $S$ limit.

Lastly, we study a spin-1 dimer compound as an illustration of the
quantum complementarity relation. In experiments, the compound exhibits
low-temperature magnetization plateaus. Our theoretical calculations
reproduce these plateaus and further show that if the system is entangled,
the magnetization plateaus coexist with the entanglement plateaus.
Successive plateaus are connected by sharp changes in the magnetization
and the amount of entanglement. The increase in one quantity is compensated
by a decrease in the other quantity so that the complementarity relation
is not violated. A large number of AFM compounds exhibit the phenomenon
of magnetization plateaus \cite{key-27}. If these systems are entangled
at the temperatures for which magnetization plateaus are observed,
one can predict the coexistence of magnetization and entanglement
plateaus in such systems. The Oshikawa, Yamanaka, Affleck (OYA) \cite{key-28}
theorem provides the condition for the occurrence of magnetization
plateaus in quasi-1d AFM systems. Magnetization plateaus have also
been observed in a two-dimensional S=$\frac{1}{2}$ AFM system $SrCu_{2}\left(BO_{3}\right)_{2}$,
thus extending the scope for the applicability of the OYA theorem.
It will be of interest to establish a connection between the OYA theorem
and the quantum complementarity relation so that the conditions for
the simultaneous appearance of the magnetization and the entanglement
plateaus are clearly identified.

\noindent \textbf{Acknowledgment.} Amit Tribedi is supported by the
Council of Scientific and Industrial Research, India under Grant No.
9/15 (306)/ 2004-EMR-I.

\appendix

\section*{Appendix A: Eigenvalues and eigenvectors of linear chain tetramer}

The Hamiltonian describing the linear chain tetramer is given in Eq.
(7). The total spin of the tetramer is $S^{tot}$. The first index
in the subscript of an eigenvector refers to the eigenvalue and the
second to $S_{z}^{tot}$, the z-component of the total spin.

$S^{tot}=2$ :

\[
\begin{array}{c}
E_{1}=\left(\frac{1}{2}+\frac{\alpha}{4}\right)J\end{array}\qquad(A1)\]

\[
\begin{array}{c}
\psi_{1,2}=\left|\uparrow\uparrow\uparrow\uparrow\right\rangle \\
\psi_{1,1}=\frac{1}{2}\left(\left|\downarrow\uparrow\uparrow\uparrow\right\rangle +\left|\uparrow\downarrow\uparrow\uparrow\right\rangle +\left|\uparrow\uparrow\downarrow\uparrow\right\rangle +\left|\uparrow\uparrow\uparrow\downarrow\right\rangle \right)\\
\psi_{1,0}=\frac{1}{\sqrt{6}}\left(\left|\uparrow\uparrow\downarrow\downarrow\right\rangle +\left|\uparrow\downarrow\uparrow\downarrow\right\rangle +\left|\uparrow\downarrow\downarrow\uparrow\right\rangle +\left|\downarrow\uparrow\uparrow\downarrow\right\rangle +\left|\downarrow\uparrow\downarrow\uparrow\right\rangle +\left|\downarrow\downarrow\uparrow\uparrow\right\rangle \right)\qquad(A2)\\
\psi_{1,-1}=\frac{1}{2}\left(\left|\uparrow\downarrow\downarrow\downarrow\right\rangle +\left|\downarrow\uparrow\downarrow\downarrow\right\rangle +\left|\downarrow\downarrow\uparrow\downarrow\right\rangle +\left|\downarrow\downarrow\downarrow\uparrow\right\rangle \right)\\
\psi_{1,-2}=\left|\downarrow\downarrow\downarrow\downarrow\right\rangle \end{array}\]

$S^{tot}=1$ : \[
E_{2}=(-\frac{1}{2}+\frac{\alpha}{4})J\qquad(A3)\]

\[
\begin{array}{c}
\psi_{2,1}=\frac{1}{2}\left(\left|\downarrow\uparrow\uparrow\uparrow\right\rangle -\left|\uparrow\downarrow\uparrow\uparrow\right\rangle -\left|\uparrow\uparrow\downarrow\uparrow\right\rangle +\left|\uparrow\uparrow\uparrow\downarrow\right\rangle \right)\\
\psi_{2,-1}=\frac{1}{2}\left(\left|\downarrow\uparrow\uparrow\uparrow\right\rangle -\left|\uparrow\downarrow\uparrow\uparrow\right\rangle -\left|\uparrow\uparrow\downarrow\uparrow\right\rangle +\left|\uparrow\uparrow\uparrow\downarrow\right\rangle \right)\\
\psi_{2,0}=\frac{1}{\sqrt{6}}\left(\left|\uparrow\downarrow\downarrow\uparrow\right\rangle -\left|\downarrow\uparrow\uparrow\downarrow\right\rangle \right)\end{array}\qquad(A4)\]

\begin{flushleft}\[
E_{3}=(-\frac{\alpha}{4}+\frac{1}{2}\sqrt{1+\alpha^{2}})J\qquad(A5)\]
\end{flushleft}

\[
\begin{array}{c}
\psi_{3,1}=\frac{1}{N_{1}}\left(\left|\downarrow\uparrow\uparrow\uparrow\right\rangle -\left|\uparrow\uparrow\uparrow\downarrow\right\rangle -\left(\alpha-\sqrt{1+\alpha^{2}}\right)(\left|\uparrow\downarrow\uparrow\uparrow\right\rangle -\left|\uparrow\uparrow\downarrow\uparrow\right\rangle \right)\\
\psi_{3,-1}=\frac{1}{N_{1}}\left(\left|\uparrow\downarrow\downarrow\downarrow\right\rangle -\left|\downarrow\downarrow\downarrow\uparrow\right\rangle -\left(\alpha-\sqrt{1+\alpha^{2}}\right)(\left|\downarrow\uparrow\downarrow\downarrow\right\rangle -\left|\downarrow\downarrow\uparrow\downarrow\right\rangle \right)\qquad\\
\psi_{3,0}=\frac{1}{N_{2}}\left(\left|\uparrow\uparrow\downarrow\downarrow\right\rangle -\left|\downarrow\downarrow\uparrow\uparrow\right\rangle -\left(\frac{1}{\alpha}-\sqrt{1+\frac{1}{\alpha^{2}}}\right)(\left|\uparrow\downarrow\uparrow\downarrow\right\rangle -\left|\downarrow\uparrow\downarrow\uparrow\right\rangle \right)\end{array}(A6)\]

\[
E_{4}=(-\frac{\alpha}{4}-\frac{1}{2}\sqrt{1+\alpha^{2}})\qquad(A7)\]

\[
\begin{array}{c}
\psi_{4,1}=\frac{1}{N_{3}}\left(\left|\downarrow\uparrow\uparrow\uparrow\right\rangle -\left|\uparrow\uparrow\uparrow\downarrow\right\rangle -\left(\alpha+\sqrt{1+\alpha^{2}}\right)(\left|\uparrow\downarrow\uparrow\uparrow\right\rangle -\left|\uparrow\uparrow\downarrow\uparrow\right\rangle \right)\\
\psi_{4,-1}=\frac{1}{N_{3}}\left(\left|\uparrow\downarrow\downarrow\downarrow\right\rangle -\left|\downarrow\downarrow\downarrow\uparrow\right\rangle -\left(\alpha+\sqrt{1+\alpha^{2}}\right)(\left|\downarrow\uparrow\downarrow\downarrow\right\rangle -\left|\downarrow\downarrow\uparrow\downarrow\right\rangle \right)\qquad(A8)\\
\psi_{4,0}=\frac{1}{N_{4}}\left(\left|\uparrow\uparrow\downarrow\downarrow\right\rangle -\left|\downarrow\downarrow\uparrow\uparrow\right\rangle -\left(\frac{1}{\alpha}+\sqrt{1+\frac{1}{\alpha^{2}}}\right)(\left|\uparrow\downarrow\uparrow\downarrow\right\rangle -\left|\downarrow\uparrow\downarrow\uparrow\right\rangle \right)\end{array}\]

$S^{tot}=0$ : 

\[
E_{5}=\left\{ -\left(\frac{1}{2}+\frac{\alpha}{4}\right)+\sqrt{1-\frac{\alpha}{2}+\frac{\alpha^{2}}{4}}\right\} J\qquad(A9)\]

\[
\psi_{5,0}=\frac{1}{N_{5}}\left(a_{1}\left|\uparrow\uparrow\downarrow\downarrow\right\rangle +b_{1}\left|\uparrow\downarrow\uparrow\downarrow\right\rangle +c_{1}\left|\uparrow\downarrow\downarrow\uparrow\right\rangle +d_{1}\left|\downarrow\uparrow\uparrow\downarrow\right\rangle +e_{1}\left|\downarrow\uparrow\downarrow\uparrow\right\rangle +f_{1}\left|\downarrow\downarrow\uparrow\uparrow\right\rangle \right)\qquad(A10)\]

\[
E_{6}=\left\{ -\left(\frac{1}{2}+\frac{\alpha}{4}\right)-\sqrt{1-\frac{\alpha}{2}+\frac{\alpha^{2}}{4}}\right\} J\qquad(A11)\]

\[
\psi_{5,0}=\frac{1}{N_{6}}\left(a_{2}\left|\uparrow\uparrow\downarrow\downarrow\right\rangle +b_{2}\left|\uparrow\downarrow\uparrow\downarrow\right\rangle +c_{2}\left|\uparrow\downarrow\downarrow\uparrow\right\rangle +d_{2}\left|\downarrow\uparrow\uparrow\downarrow\right\rangle +e_{2}\left|\downarrow\uparrow\downarrow\uparrow\right\rangle +f_{2}\left|\downarrow\downarrow\uparrow\uparrow\right\rangle \right)\qquad(A12)\]

\noindent where $a_{1}=f_{1}=a_{2}=f_{2}=1,\: b_{1}=-\frac{2}{\alpha}+2\sqrt{\frac{1}{4}-\frac{1}{2\alpha}+\frac{1}{\alpha^{2}}}=e_{1},\: b_{2}=-\frac{2}{\alpha}-2\sqrt{\frac{1}{4}-\frac{1}{2\alpha}+\frac{1}{\alpha^{2}}}=e_{2}$,
$\:$$c_{1}=-1+\frac{2}{\alpha}-2\sqrt{\frac{1}{4}-\frac{1}{2\alpha}+\frac{1}{\alpha^{2}}}=d_{1},\: c_{2}=-1+\frac{2}{\alpha}+2\sqrt{\frac{1}{4}-\frac{1}{2\alpha}+\frac{1}{\alpha^{2}}}=d_{2}$

\noindent $N_{1}$, $N_{2}$, $N_{3}$, $N_{4}$, $N_{5}$ and $N_{6}$
are the appropriate normalization constants.

\section*{Appendix B: Eigenvalues and eigenvectors of the spin-1 dimer}

The dimer Hamiltonian $H_{d}$ is given by Eq. (21). The basis functions
are represented as $\left|S_{1}^{z},S_{2}^{z}\right\rangle $ with
$S_{1}^{z}=\pm1,0$ and $S_{2}^{z}=\pm1,0$. The eigenstates and the
eigenvalues are described by $\phi_{n,m}$ and $\lambda_{n,m}$ where
$`n'$ and $`m'$ refer to the total spin $S^{tot}$ of the dimer
and its z-component respectively.

$S^{tot}=2$ : \[
\begin{array}{c}
\phi_{2,\pm2}=\left|\pm1,\pm1\right\rangle \\
\lambda_{2,\pm2}=\delta J+2d\pm2B\end{array}\qquad(B1)\]

\[
\begin{array}{c}
\phi_{2,\pm1}=\frac{1}{\sqrt{2}}\left(\left|\pm1,0\right\rangle +\left|0,\pm1\right\rangle \right)\\
\lambda_{2,\pm1}=J+d\pm B\end{array}\qquad(B2)\]

\[
\begin{array}{c}
\phi_{2,0}=\frac{1}{2}\left(A_{m}\left|1,-1\right\rangle +\left|-1,1\right\rangle \right)+\sqrt{2}A_{p}\left|0,0\right\rangle \\
\lambda_{2,0}=-\frac{\delta J}{2}+d+R\end{array}\qquad(B3)\]

$S^{tot}=1$ : \[
\begin{array}{c}
\phi_{1,\pm1}=\frac{1}{\sqrt{2}}\left(\left|\pm1,0\right\rangle -\left|0,\pm1\right\rangle \right)\\
\lambda_{1,\pm1}=-J+d\pm B\end{array}\qquad(B4)\]

\[
\begin{array}{c}
\phi_{1,0}=\frac{1}{\sqrt{2}}\left(\left|1,-1\right\rangle -\left|-1,1\right\rangle \right)\\
\lambda_{2,\pm1}=-\delta J+2d\end{array}\qquad(B5)\]

$S^{tot}=0$ : \[
\begin{array}{c}
\phi_{0,0}=\frac{1}{2}\left(A_{p}\left(\left|1,-1\right\rangle +\left|-1,1\right\rangle \right)-\sqrt{2}A_{m}\left|0,0\right\rangle \right)\\
\lambda_{0,0}=-\frac{\delta J}{2}+d-R\end{array}\qquad(B6)\]

where\[
\begin{array}{c}
R=\left[\left(\frac{\delta J}{2}-d\right)^{2}+2J^{2}\right]^{\frac{1}{2}}\\
A_{p}=\left(\frac{R+\frac{\delta J}{2}-d}{R}\right)^{\frac{1}{2}}\\
A_{m}=\left(\frac{R-\frac{\delta J}{2}+d}{R}\right)^{\frac{1}{2}}\end{array}\qquad(B7)\]

\end{document}